\documentclass[useAMS,usenatbib]{mn2e}

\usepackage{psfig}
\usepackage{epsfig,afterpage}
\usepackage{graphicx}
\usepackage{pifont}

\title[]{Automated derivation of stellar atmospheric parameters and chemical abundances: the MATISSE algorithm.} 
\author[A. Recio-Blanco, A. Bijaoui and P. de Laverny]{A. Recio-Blanco$^{1}$, A. Bijaoui$^{1}$ and P. de Laverny$^{1}$ 
\thanks{E-mail: arecio@obs-nice.fr, bijaoui@obs-nice.fr, laverny@obs-nice.fr 
}\\ 
$^{1}$Dpt. Cassiop\'ee, UMR 6202, Observatoire de la C\^ote d'Azur, B.P.4229, F-06304 Nice Cedex 4, France\\} 
\begin{document} 
 
\date{Accepted Year Month day.  Received 2005 Month day; in original form 2005 Month day} 
 
\pagerange{\pageref{firstpage}--\pageref{lastpage}} \pubyear{2002} 
 
\maketitle 
 
\label{firstpage} 
 
\begin{abstract} 
We present an automated procedure for the derivation of  atmospheric
parameters (Teff, log~$g$, [M/H]) and individual chemical abundances from stellar
spectra. The MATrix Inversion for Spectral SythEsis (MATISSE) algorithm
determines a basis, {\it B}$_\theta$($\lambda$),
allowing to derive a particular stellar parameter  $\theta$ 
by projection of an observed spectrum.  
The {\it B}$_\theta$($\lambda$) function is determined from an optimal linear combination of 
theoretical spectra and it relates,  in a quantitative way, the variations in the spectrum flux
with variations in $\theta$.
An application of this method to the GAIA/RVS spectral
range is described, together  with its performances for different types of stars of various
metallicities. Blind tests with synthetic spectra of randomly selected
parameters and observed input spectra are also presented. 
The method gives rapid, accurate and stable results  and it can be efficiently applied to
the study of stellar populations through the analysis of large spectral data sets, including moderate
to low signal to noise spectra.
\end{abstract} 
 
\begin{keywords} 
stars: abundances - stars: fundamental parameters - methods: data analysis - techniques: spectroscopy -
Galaxy: stellar content
\end{keywords} 
 
\section{Introduction} 
 
The physical parametrization of stars is a crucial step 
in our comprehension of stellar and galactic astrophysics. 
By way of illustration, it could be cited the impact of the 
Hertzsprung-Russel (HR) diagram on our understanding of stellar 
evolution, or the identification and characterization of
the different components and stellar populations of the Milky
Way and its satellites. The classification of stars, from rich collections of 
stellar properties and vast catalogues of objects, is now opening
new horizons in, for instance, the disentangling of the detailed sequence 
of events which led to the Milky Way galaxy. 
In this sense, the chemical parametrization of stars is mandatory for a
fruitful classification attempt, reinforcing the traditional approach. 
The chemical abundance ratios of some
species to anothers (e.g. the [$\alpha$/Fe] ratio) provide an indication of 
the star formation history time scale
and details of the chemical enrichment for one specific population.

In the recent years, the use of spectrographs with multiobject 
capabilities (FLAMES, 2dF...) and the implementation of extensive surveys (Sloan, RAVE,...)
has enormously increased the
quantity of available data and, as a result, 
the efforts needed for their analysis. In the near future, the European Space Agency Gaia mission will collect 
several millions of stellar spectra, allowing a very complete mapping of the 
Milky Way.  Moreover, thanks to the spectral resolution (R$=$11\,500) of the Gaia Radial Velocity Spectrograph 
(RVS), it will be possible to derive individual chemical element abundances,
leading to a huge and precious database. In this context, automated techniques of spectral analysis
and  classification are  needed, in order to perform a rapid and homogeneous
processing of the data and to allow an efficient scientific return.

The intrinsic physical properties of stars, can be
indirectly derived from stellar spectra. Effective
temperature ($T_{\rm{eff}}$), surface gravity (log~$g$), and chemical composition
are the main parameters governing the stellar atmosphere
and non-linear combinations of them characterize the
features on a stellar spectrum.
The knowledge of these quantities, together with
stellar structure and evolution models, can help to
determine a star's history and future evolution. 


Traditionally and for historical reasons, 
the more empirical variables of luminosity
and colour index  have been used to indirectly
describe spectroscopic parameters.
The MK system (Morgan, Keenan \& Kellman, 1943) avoids the  
complexities of spectral lines formation by the use of  
standard spectra. Spectral class and luminosity type are
assigned from medium-low 
resolution spectra in, essentially, a two-parameters  
classification effort, representative of $T_{\rm{eff}}$ and log~$g$. Nevertheless, this approach suffers from a
series of disadvantages, mainly arising from the
lack of a metallicity quantification. The almost neglect of the chemical dimension 
is the cause of an important loss of scientific information.
Moreover, since the match between the object spectrum and the MK standards is visually
performed, the classification suffers from subjective decisions.

 
On the other hand, a classification directly based on 
the physical parameters, although model dependent,
takes more naturally into account the scientific
quantities of interest and
offers a continuous parametrization instead of a
discrete spectral/luminosity type assignment.


We describe here the MATrix Inversion for Spectral SythEsis (MATISSE) algorithm.
It represents a new effort in automatic spectral analysis
(for discussion of already developed methods see Bailer-Jones, 2001).
The parametrization problem, applicable to millions of stellar spectra, is approached
in a way that tries to tackle some disadvantages of already existing automated
classification techniques, as excessive computing times.
This method uses the inversion of the covariance 
matrix of a grid of synthetic spectra to determine a basis allowing to derive a
particular stellar parameter by projection of an object spectrum.
An initial attempt was reported by Th\'evenin, Bijaoui \& Katz (2002). 
 
 The motivation and the mathematical description of the method are developed in Section 2.
Section 3 presents all the steps in the application of MATISSE 
and the results on the measurement of stellar atmospheric parameters ( $T_{\rm{eff}}$, log~$g$, global metallicity [M/H]) and
[$\alpha$/Fe] abundance for stars in different parts of the Hertzsprung-Russel 
diagram. The procedure performances are illustrated  for input data in the Gaia/RVS
wavelength domain, including real spectra. 
Final discussions and conclusions are presented in Section~4.

\section[]{Mathematical base of the MATISSE algorithm} 
 
Let us consider an observed spectrum $O(\lambda)$, corrupted by a
Gaussian noise, independent of $\lambda$, of standard deviation $\sigma$.
A grid of theoretical spectra $S(\lambda,\theta)$, where $\theta$
corresponds to the stellar parameters, is implemented for
its analysis. 
In this framework, the parameters estimation problem consists into finding the minimum distance
$d(\theta)$ between $O(\lambda)$ and the spectra $S(\lambda,\theta$). It is a
typical least mean square estimation problem. Its
classical solution is obtained by solving the normal equations
resulting from the $d(\theta)$ derivation as a function of each parameter.
Nevertheless, many problems arise from this approach:
\begin{itemize}
  \item Stellar spectra do not depend linearly on the physical
  parameters, so that an iterative procedure must be done, using a
  local linearization.
  \item The synthetic spectra are previously computed on a set of discrete
  parameters. The iterative procedure needs to compute many models
  with enough resolution  for the targeted precision.
  Generally, this operation can not be done. The estimation can be only
  obtained from a set of given models.
  \item The distance is not necessarily a convex function of the 
parameters. So the
  gradient descent can lead to a local minimum, far from the
  correct one.
\end{itemize}

As a consequence, instead of solving the normal equations after linearization, it is
better, for a given set of models, to find directly the minimum by
i) computing all the distances to the models and ii) interpolating 
between the distances corresponding to the neighbour parameter values 
(limiting the computations to a restricted cell of models).
This approach needs
the computation of a theoretical grid with the enough resolution
to get a convex function for the distance, in the volume restricted to each cell. 
Within a cell, the spectral
distribution should be $quasi$ linear as a function of the parameters.
This insurance can not be obtained by
theoretical derivations but by experimentation.  In
this situation, the determination of the set of parameters,
corresponding to the minimum distance, leads to a solution with
the accuracy of the grid resolution. Therefore, the problem 
consists into increasing the accuracy without computing new models.

Already existent automatic classification methods, mainly minimum
distances techniques, genetic algorithms and neural networks, try
to tackle the above described problem in different ways (see also
examples in Sect. 4), none of them without disadvantages.

As mentioned before, pure minimum distances methods are limited
by the theoretical spectra grid resolution. Moreover, 
border effects with respect to the selected
grid can be considerable. The time for the data processing is also heavy, 
when dealing with a grid of several thousand elements.
Genetic algorithms can also suffer from excessive computing times,
critical dependences on the algorithm parametrization and difficulties
in evaluating the suitability of the final convergence.
Finally, in neural networks methods, the individual relations between 
the input variables and the output variables have not an analytical basis. 
As a consequence, the impact of the underlying physical laws and parameters is
more difficult to estimate. In addition, they need a hierarchical structure
that is unknown $a ~priori$.

In this context, the method presented here opens an unexplored
pathway. Two approaches were initially examined, in order to solve
the interpolation problem described above:

\begin{itemize}
  \item The objective analysis (Recio-Blanco et al., 2005), which determines 
a parameter through a weighted combination of the corresponding cell values. The weights are
derived from the distances between the observed spectrum and the 
models. Exponential weights gave the best
results. Nevertheless, the results were not very satisfying. In 
particular, a bias appeared for values near the limits of the grid, as for
classical minimum distances techniques.
  \item An algorithm based on the projection of the observed spectrum on 
specific
basic vectors. These vectors are a combination of the neighbour
models. This algorithm, MATISSE, carried out to the best results and is
described below.
\end{itemize}

The idea of the MATISSE algorithm came from the classical
Principal Component Analysis (PCA). From the PCA we get a set of
decorrelated components as combinations of the measurements (in
our problem the spectral intensities). The weights of the
combinations are the eigenvectors of the variance-covariance
matrix. One could imagine that the principal components would correspond
also to physical parameters. The experiments showed that it was
not true. So, the question was to derive the weights which gives
coefficients as close as possible to the physical parameters. The
criterion we adopted  was the statistical correlation between the
input and the output parameter values.

In other words, the implemented algorithm determines a vector, {\it B}$_\theta$($\lambda$),
allowing to derive a particular stellar parameter  $\theta$ 
by projection of an input spectrum on it.  This $\theta$  parameter
can be the effective temperature, the gravity, the global metallicity, the [$\alpha$/Fe]
content, individual chemical abundances, v~sin{\it i} value, etc...
The {\it B}$_\theta$($\lambda$) function is derived from an optimal linear combination of 
theoretical spectra and it relates,  in a quantitative way, the variations in the spectrum flux
with the $\theta$ variations.


First of all, the data on a particular $\theta$ variable
and the spectra of the grid are 
subtracted of their mean value.
The {\it B}$_\theta$($\lambda$) basis is then constructed from a linear combination of spectra, with
$\alpha_i$ being the weight associated to the spectrum S$_i(\lambda)$:

\begin{equation}
B_\theta(\lambda) = \sum_{i} ~\alpha_i  S_i(\lambda)
\end{equation}

The parameter $\theta_i$ is estimated by the projection 
of a spectrum into the corresponding basis vector:

\begin{equation}
\hat {\theta_i}  = \sum_{ \lambda } ~B_\theta(\lambda) S_i(\lambda)
\end{equation}


\noindent with $\hat {\theta_i}$ being the recovered value.


\smallskip

Combining Eq. 1 and Eq. 2 we obtain :

\begin{equation}
\hat {\theta_i}  = \sum_{j} ~c_{ij} \alpha_j
\end{equation}

where $c_{ij}$ is the correlation value between the spectra S$_i$
and S$_j$. Taking into account that the spectra have been subtracted of
their mean value, $c_{ij}$ can be interpreted as the covariance between the 
S$_i$ and S$_j$, if the spectral values are considerered as random variables.


The $\alpha_i$ are obtained from the maximum correlation between
$\theta_i$ and $\hat\theta_i$. Therefore, the following relation:

\begin{equation}
R = \frac {( \sum_{i} \hat {\theta_i} \theta_i )^2}{\sum_{i} \hat {\theta_i}^2} 
\end{equation}

\smallskip

has to be maximized.
Using Eq. 3, the above expression can be written in the form:

\begin{equation}
R = \frac {( \sum_j \alpha_j (\sum_i  c_{ij} \theta_i))^2}{\sum_{j,k} ( \alpha_j \alpha_k (\sum_i c_{ij} c_{ik}) )}
\end{equation}

Hence, maximizing $R$ we obtain:

\begin{equation}
\sum_k (\sum_i c_{ij} c_{ik}) \alpha_k = a ~ (\sum_i  c_{ij} \theta_i)
\end{equation}

with {\it a}  being a scalar quantity that we can impose to be equal
to 1. Therefore, for an invertible covariance matrix, we would have:

\begin{equation}
\sum_k c_{ik} \alpha_k = \theta_i
\end{equation}

In our case, the covariance matrix is empirically found to be non invertible.
As a consequence, Eq. (6) has to be solved through a least squares linear regression between
$\theta$ and $\hat \theta$. 
The optimal $\alpha_i$ values can be calculated, for instance, from an iterative Van Cittert's 
algorithm (Van Cittert, 1931).
%
%

In other words, we
determine a linear relation between the values of a parameter $\theta_i$ for
a given set of spectra, S$_i$($\lambda$), and the product  S$_i$($\lambda$) 
{\it B}$_\theta$($\lambda$). 

Then, to determine the parameters of an object spectrum,
not belonging to the learning set, it is only necessary to multiply it
by the corresponding {\it B}$_\theta$($\lambda$) and to transform the result using the derived
linear regression between $\hat \theta$ and $\theta$. This procedure is therefore
extremely rapid and ideal for the analysis of huge quantities of data.

In order to avoid the effects of important non-linear variations
in the spectra, it is adviceable to restrict the working domain to a subregion
of the spectra grid. To this purpose, a two level procedure can be followed,
by deriving initial {\it B}$^{\rm{o}}_\theta$($\lambda$) functions to make
a preliminary guess in the parameters and then to refine the result using
local {\it B}$^{~l}_\theta$($\lambda$) functions (see Section 3.2). This initial
guess can also be done through the evaluation of available photometry for the targets.

\section[]{Application of MATISSE to spectra in the Gaia RVS domain}


In order to illustrate the performances of the algorithm, this Section describes
its application to the measurement of stellar atmospheric parameters ($T_{\rm{eff}}$, log~$g$, [M/H]) and
[$\alpha$/Fe] abundance for stars in different parts of the Hertzsprung-Russel
diagram. The Gaia/RVS spectral domain has been selected as the framework for this example,
due to its large potential in the analysis of stellar spectra  for Galactic studies 
(Munari, 2003 and the RAVE consortia studies). It ranges from $\lambda$ 8470-8740 \AA, containing 
the infrared triplet of ionized calcium.
Thanks to the RVS spectral resolution, 
R$=$11\,500, several atomic and molecular lines,
depending on stellar spectral type and signal to noise ratio, can be identified.
The most relevant features in the RVS wavelength range
are the ionized calcium triplet lines, together with the hydrogen Paschen lines.
Many lines of iron and iron peak elements can also be found. Moreover,
besides of calcium, other $\alpha$-elements like silicium, sulfur, magnesium and
titanium have features in the RVS range. One s-process element, the zirconium,
has one measurable line in the appropriate conditions of signal to  noise and
atmospheric parameters. Finally, molecular lines, like the TiO and CN
lines, are detectable for the coolest stars.

All the steps in the application of MATISSE are described below, including the used
synthetic spectra grid. The procedure performances are detailed through the results
for trial spectra and blind tests with unknown input flux calibrated (Sect. 3.3) and
normalized (Sect. 3.4) spectra.

\subsection[]{The grid of synthetic stellar spectra}  
 
The computed grid of theoretical stellar spectra covering the Gaia/RVS
domain (see Recio-Blanco, de Laverny \& Plez, 2005, for a more detailed description) 
is based on a new generation 
of MARCS model atmospheres and mostly devoted to FGK dwarf and giant stars.
It was computed with the  turbospectrum code
(Alvarez \& Plez 1998, and further improvements by  Plez)
in plane-parallel and spherical geometry (depending on the gravity).

The adopted line list consists in all the atomic lines found in the
VALD database (Kupka et al., 1999) and molecular lines.
The molecular line list includes ZrO, TiO, VO,
CN,  C$_2$, CH, SiH, CaH and MgH lines with their
corresponding   isotopic   variations. Collisional broadening by atomic hydrogen 
of several atomic lines are computed
as in Barklem et al. (2000).
Other lines are broadenned with the classical theory (see Recio-Blanco, 
de Laverny \& Plez, 2005, for further details about these linelists).
We note that the treatment of the line broadening is 
an important issue when comparing synthetic to observed spectra and that could introduce
some biases in the derived atmospheric parameters. However, this article
is focussed on a new algorithm for the derivation of such parameters and
its internal validity is actually independent of that issue.


Regarding the model atmospheres, we used  a new grid of MARCS
(version of November 2004)
 one-dimensional, plane-parallel and spherical LTE model atmospheres
(Gustafsson et al., 2003; Gustafsson et al., 2006). Turbulence pressure was
included and convection was simulated according to the local
mixing-length recipe.
The grid consists in 1\,858 stellar models with
effective temperatures between 4\,000~K and
8\,000~K (step 250~K), logarithmic surface gravities between -1.0 to 5.0
(step 0.5), and overall metallicities between -5.0 and 1.0 (with a variable
step from 1.0 to 0.25 dex) and different $\alpha-$element enhancements (see below).
The adopted solar abundances for C, N and O in the atmospheric models are from
 the recent revision by Asplund et al. (2005).
Solar abundances of other chemical species are from Grevesse 
\& Sauval (1998).

From this grid of model atmospheres, synthetic spectra have been
computed in the range $\lambda=8\,475$\AA \, to $\lambda=8\,745$\AA \,
with a step of 0.02\AA  ~and then convolved and re-sampled
to match the RVS spectral resolution
(R~$\sim$~11\,500). The number of pixels per resolution element is 3,
and for historical reasons, slightly different from the one presently 
defined for Gaia (2 pixels for resolution element).

The synthetic spectra have been computed with the same
geometry and abundances as in the model atmosphere.
Plane-parallel models have been used for log~$g$
from +3.0 to +5.0. They have a microturbulent-velocity parameter of 1.0~km/s.
Models with spherical geometry have been considered for giant stars with
log~$g < $+3.0. These models have been computed for masses of 1.0~M$_\odot$ and
with a microturbulence parameter of 2.0~km/s. For metal-poor models, the following
$\alpha$-enhancements were considered:
[$\alpha$/Fe]=+0.1 for [Fe/H]=-0.25, [$\alpha$/Fe]=+0.2 for [Fe/H]=-0.5,
[$\alpha$/Fe]=+0.3 for [Fe/H]=-0.75
and [$\alpha$/Fe]=+0.4 for [Fe/H]$\leq$-1.00.
Chemical species treated as $\alpha$-elements are Ne, Mg, Si, S, Ar,
Ca and Ti.
Oxygen also follows the same enhancements.

For each model atmosphere, we furthermore considered
$\alpha$-elements abundance variations of +0.4, +0.2, +0.0,
-0.2, -0.4~dex, with respect to the original abundances in the model.
Five synthetic spectra have thus been computed
per model atmosphere leading to a final library of 9\,290 spectra.

On the other hand, we point out that this grid of spectra
is not optimized for all types of stars. In particular, we did not check the
validity of all the considered line data and the classical assumptions (LTE, hydrostatic), used
when deriving the atmospheric structure, can be questioned.
However, a comparison of some synthetic spectra with observed ones shows a
rather good match in the RVS domain. Finally, for the purpose of this work, the
use of high quality synthetic spectra (still not available for almost any
region in the Hertzsprung-Russel diagram) is not imperative, since we focus on the
algorithm itself. 

\begin{figure*}
\centering
\includegraphics[width=17cm]{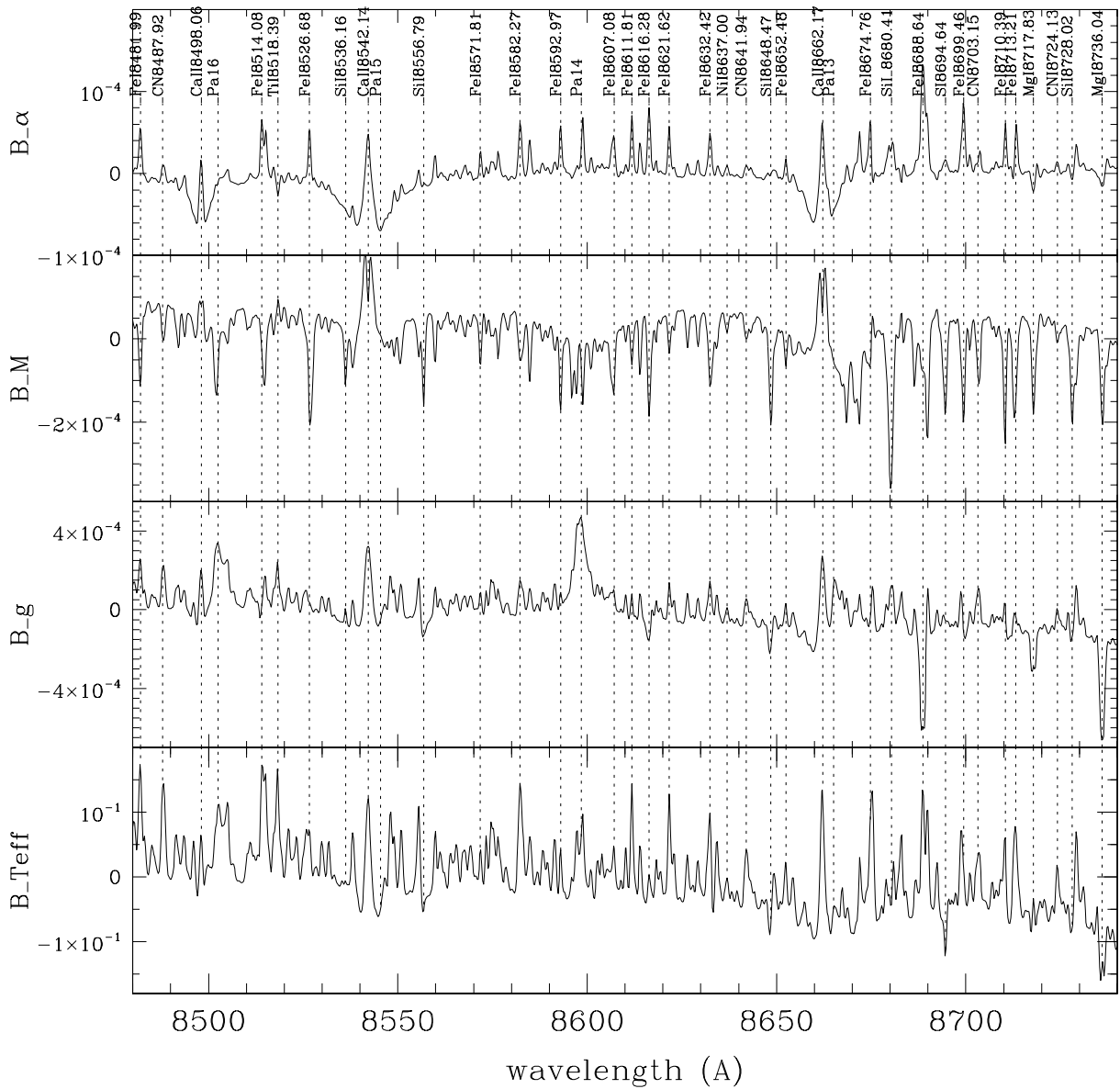}\\
\caption{{\it B} functions for the effective temperature, the gravity, the
global metallicity and the [$\alpha$/Fe] content of a solar type star (the units are related to
the considered parameter ones, per flux unit).
The spectral lines identification in the upper panel allows 
to see how a particular {\it B}$_\theta$($\lambda$) function deviates more from zero at the wavelengths
containing the highest quantity of information on the corresponding parameter $\theta$.  }
\label{fig1}
\end{figure*}

\subsection[]{Derivation of the {\it B}$_\theta$($\lambda$) functions}

One of the advantages of the MATISSE algorithm with respect to other automatic
spectral parameters derivation procedures is the fact that the calculated 
 {\it B}$_\theta$($\lambda$) functions allow the control of the spectral features
involved in the analysis. 
Basically, {\it B}$_\theta$($\lambda$)
deviates more from zero at the wavelengths mostly affected by a
change in  $\theta$, that is, at the spectral regions containing the highest
quantity of information on a given parameter for the stellar types considered
in the spectra grid. In this way, we are informed on
which lines are been used for each parameter under consideration, in a more
physical approach close to the traditional spectral synthesis. Similarly,
if we prefer to avoid the use of a certain line due, for instance, to problems 
of unreliable atomic data, we can exclude the corresponding wavelength region
from the algorithm that calculates the {\it B}$_\theta$($\lambda$) functions.

As a way of illustration, Figure 1 shows the derived {\it B}$_\theta$($\lambda$) functions
for a solar type star, corresponding to  the determination of effective
temperature, {\it B}$_{T_{\rm eff}}$($\lambda$), surface gravity, {\it B}$_g$($\lambda$), global
metallicity, {\it B}$_M$($\lambda$) and $\alpha$-element abundance with respect to iron, 
{\it B}$_\alpha$($\lambda$). 
These functions have been calculated for a 
subgrid of synthetic spectra around the point with 
$T_{\rm{eff}} = 5750$~K, log~$g = 4.5$~dex, [M/H]$ = 0.0$~dex, [$\alpha$/Fe]$ = 0.0$~dex.
The parameter dimensions of the grid subregion are 1000 K (for {\it B}$_{T_{\rm eff}}$($\lambda$)) or
500 K (for the other three functions), 1.0 dex, 1.0 dex and
0.4 dex respectively. 
The most important lines in a solar type spectrum are also marked in Fig. 1. 

First of all, it should be noted that the most relevant features
of each {\it B}$_\theta$($\lambda$) correspond to 
typical lines of the stars under analysis: Paschen lines, CaII triplet,
Fe, Si, Mg lines,... 
In addition, {\it B}$_{T_{\rm eff}}$($\lambda$) has a slope, positive at lower wavelengths,
that reflects the temperature dependence of the continuum flux. Several
metallic lines are also good temperature indicators: FeI lines (e.g.
8484.99 \AA, 8514.08 \AA, 8582.27 \AA, 8611.81 \AA) and TiI (8518.40 \AA)
with increasing intensity at higher $T_{\rm{eff}}$; MgI (8736.04 \AA) and
SiI (8694.64 \AA) with an inverse behaviour. Of the ionized calcium triplet
lines, the two reddest ones (8542.14 \AA  ~and 8662.17 \AA), including 
their wings, are those containing more temperature information for this 
kind of stars. Finally, the Paschen lines also contribute to the temperature
evaluation of a solar type spectrum.

The most important features in the {\it B}$_g$($\lambda$) function, correspond
to the Pachen lines (specially Pa~14 at 8598.40 \AA), the CaII lines, the
neutral magnesium lines and some iron lines like FeI 8688.64 \AA. Finally,
the iron lines, together with $\alpha$-element lines (the CaII triplet, magnesium and 
silicium) dominate the information on the stellar global metallicity and 
the $\alpha$-element abundance. 

The MATISSE algorithm performs, therefore, a sort of automated 
spectral synthesis, by using the more adapted lines for the
derivation of each stellar parameter or chemical data.

On the other hand, it is worth noticing that, in order to better separate
the influence of each parameter on the spectral lines, a {\it B}$_{\theta}$($\lambda$) function
has to be built from a set of synthetic spectra sharing at least ($D-1$) parameters,
where $D$ is the dimension of the grid. In other words, once the centre of the
grid (or subgrid) we want to analyse is selected, only those spectra having ($D-1$) parameters equal
to the central point have to be used. In practice, this allows to deal with
spectra at varing $T_{\rm{eff}}$
but constant log~$g$,  [M/H] and [$\alpha$/Fe], varing log~$g$ but constant $T_{\rm{eff}}$, 
[M/H] and [$\alpha$/Fe], etc... As a consequence, it is possible to better distinguish between a
change in the spectrum caused by a change in, for instance,  $T_{\rm{eff}}$, from the effect due to
the variation of a different parameter.

\subsection[]{Application to flux calibrated spectra.}

To illustrate the performancies of MATISSE as applied to the Gaia/RVS 
spectral domain and resolution, the results for trial spectra from 
different regions of the Hertzsprung-Russel diagram are presented in the following.
In this Section, the performancies on flux calibrated spectra are described.
Actually, the flux calibration of the Gaia/RVS spectra will be allowed by
one  medium band photometric filter covering
the RVS spectral range and by accurate absolute distance measurements (from
parallaxes) for all the stars observed by the RVS.

In order to evaluate the performancies of
the method and to ensure the absence of biases and the validity of a linear treatment
as the MATISSE one, the algorithm has been applied to synthetic spectra of different
stellar types. This allows to confirm the stability of the
method at  different spectral classes, luminosity types and metallicity contents, but also
to illustrate the dependencies of the obtained performancies on the stellar atmospheric 
properties.  In particular, possible tracers of the Galactic Thin disc, Thick disc and Halo
stellar populations have been selected to this purpose: metal rich cool dwarfs 
($T_{\rm{eff}}=$5000-6000~K, log~$g >$ 3.5~dex, [M/H] $>$ -0.5~dex) intermediate
metallicity cool giants ($T_{\rm{eff}}<$5000~K, log~$g =$1.0-3.5~dex, 0.0~dex $>$ [M/H] $>$ -1.0~dex) 
and very metal poor hot subgiants ($T_{\rm{eff}}>$6000~K, log~$g =$ 2.0-4.0~dex, -1.0 $>$ [M/H] $>$ -2.5~dex).
The various [$\alpha$/Fe] contents included in the grid have been considered.

A total of 100 different cases, with randomly selected parameters, in the above ranges, have been analysed
to check if the performance is the same regardless how many parameters are initialized
from a position off the subgrid centre. In addition, to quantify the  dependence of the 
results on the quality of the input spectra, a programme
to introduce Gaussian white noise, at desiderable values of signal to noise ratio (S/N), 
was implemented. 
For each of the 100 {\it object} synthetic spectra, five signal to noise values have been considered
and  1000 noised spectra were computed for each S/N value. The MATISSE algorithm 
is then applied
to a total of 500~000 spectra, in order to derive their atmospheric parameters
and [$\alpha$/Fe] abundance.

At this stage, the use of input synthetic spectra instead of a library of real data 
guarantees the absence of additional errors that could blur the true
performances of the method (see for example Bailer-Jones, 2000 and Willemsen et al. 2003 
for other works on automated spectral analysis using synthetic spectra). 
Actually, imperfections in the physical assumptions (LTE, hydrostatic, see for instance
Asplund 2005), uncertainties in the line 
data used for the theoretical spectra computation, uncertainties in the parameters of
the real stars, etc... could produce biases on the results and complicate the
evaluation of the MATISSE procedure. In addition, the use of synthetic spectra
allows the availability of a larger number of data over a large parameter space.

First of all, a preliminary evaluation of the spectrum parameters can be done 
using MATISSE initial
{\it B}$^{\rm{o}}_\theta$($\lambda$) functions.
This permits to restrict the working domain to a subregion of the spectra grid  and to minimize the
effects of important non linear variations in the spectra, through
the considered parameter range. The preliminary values of the parameters 
can then be refined or corrected by using {\it local} {\it B}$^{~l}_\theta$($\lambda$) functions.

The {\it B}$^{\rm{o}}_\theta$($\lambda$) functions are derived from 
theoretical spectra whose parameters span 
the whole range of the grid or an important fraction of it. Therefore, as the spectral variations
as a function of the parameters 
are non-linear, the application of these {\it B}$^{\rm{o}}_\theta$($\lambda$) functions will 
suffer from biases of variable importance, depending on the covered parameter range.
On the other hand, a crucial characteristic of the stellar spectra grid is that non-linearity
is almost negligible at a local scale of about 1000~K in $T_{\rm{eff}}$, 1~dex in log~$g$ and 
1~dex in [M/H]. Thus, if the errors introduced by the
initial {\it B}$^{\rm{o}}_\theta$($\lambda$) functions are equal to or smaller than the above mentioned
limits, it is possible to converge to the correct solution. This is, in fact, the case for
the Gaia/RVS spectral domain and resolution (even if it contains rather few lines compared
to other spectral regions more in the blue). 
Table 1 presents the corresponding   
errors coming from the application of the initial 
{\it B}$^{\rm{o}}_\theta$($\lambda$) functions for flux calibrated
spectra. Those values have 
been derived for the whole set of trial spectra and additionally
verified for all the spectra of the grid with [M/H]$>$-3.0~dex. The first
guess in the [$\alpha$/Fe] is chosen as a function of the global metallicity
value, following the same  $\alpha$-enhancements considered in Section 3.1.

The achieved precisions in the atmospheric parameters 
are clearly adequate for a subgrid selection and a subsequent
refinement of the results using local {\it B}$^{~l}_\theta$($\lambda$) functions. In addition,
those values are practically the same at any signal to noise ratio, as the errors are
largely dominated by the bias, which is due to non linearity problems and independent of the noise. 

On the other hand, the number of initial {\it B}$^{\rm{o}}_\theta$($\lambda$) functions can
depend on the selected parameter and the values covered by the grid. 
In the case of the Gaia/RVS grid described in Section 3.1, only one 
{\it B}$^{\rm{o}}_{T_{\rm eff}}$($\lambda$) has been
necessary to attain the objectives of accuracy for the selection of the local
{\it almost linear} subgrids and the subsequent application of the final
{\it B}$^{~l}_{T_{\rm eff}}$($\lambda$). On the contrary, the first guesses in
gravity and metallicity are coupled to the preliminary estimation of temperature:
four different {\it B}$^{\rm{o}}_g$($\lambda$) functions and three {\it B}$^{\rm{o}}_M$($\lambda$)
functions were calculated, at different temperature intervals. The use of a
rough temperature estimation in the preliminary evaluation of the other two 
atmospheric parameters tackles the problem of degeneracy between them and reduces
the final set of initial {\it B}$^{\rm{o}}_\theta$($\lambda$) functions to eight.

One possible alternative approach to the use of the initial 
{\it B}$^{\rm{o}}_\theta$($\lambda$) functions could be the photometric
estimation of a star's atmospheric
parameters through available colours or the location on a colour-magnitude 
diagram. 

\begin{table}
\caption{Preliminary estimation of the parameters through initial
MATISSE {\it B}$^{\rm{o}}_\theta$($\lambda$) functions for the Gaia/RVS spectral domain
and resolution. They correspond to spectra with [M/H] $>$ -3 and S/N$>$25. 
The {\it B}$^{\rm{o}}_\theta$($\lambda$) functions are applied for a subgrid selection before the use
of the final local {\it B}$^{~l}_\theta$($\lambda$) functions.}
\leavevmode
 \begin{center}
           \begin{tabular}[h]{lccc}
\hline 
\noalign{\smallskip} 
 & $T_{\rm{eff}}<4500\rm{K}$ & $ 4500< T_{\rm{eff}}<6000\rm{K}$&$T_{\rm{eff}}>6000\rm{K}$ \\
\noalign{\smallskip} 
\hline 
$\Delta T_{\rm{eff}}$  & 300-600~K   & 200-600~K   & 400-750~K   \\
$\Delta \rm{log}~g$    & 0.1-1.0~dex & 0.1-0.3~dex & 0.1-0.2~dex \\
$\Delta \rm{[M/H]}$         & 0.1-0.5~dex & 0.1-0.6~dex & 0.2-0.6~dex \\
\hline 
       \end{tabular}
  \end{center}

\end{table}

\begin{figure*}
\centering
\includegraphics[width=17cm]{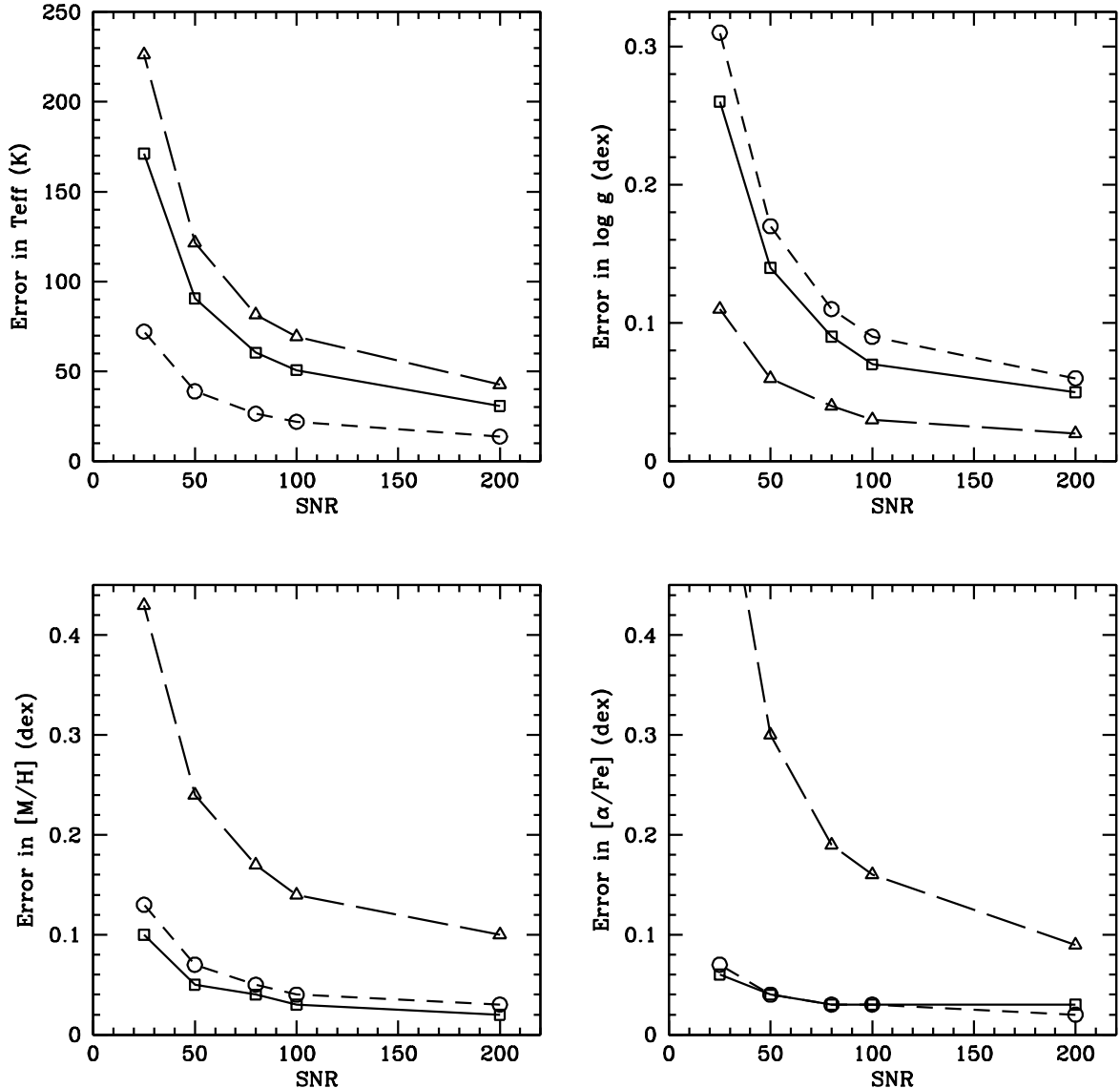}\\
\caption{Mean maximum errors in the recovered parameters as a function of the S/N, for synthetic 
flux calibrated spectra in the Gaia/RVS domain. The three different lines on
each panel correspond to metal rich cool dwarfs (solid line and
squares), intermediate metallicity cool giants 
(short dashed line and circles), metal poor hot subgiants (long dashed
line and triangles).}
\label{fig1}
\end{figure*}

\begin{figure*}
\centering
\includegraphics[width=17cm]{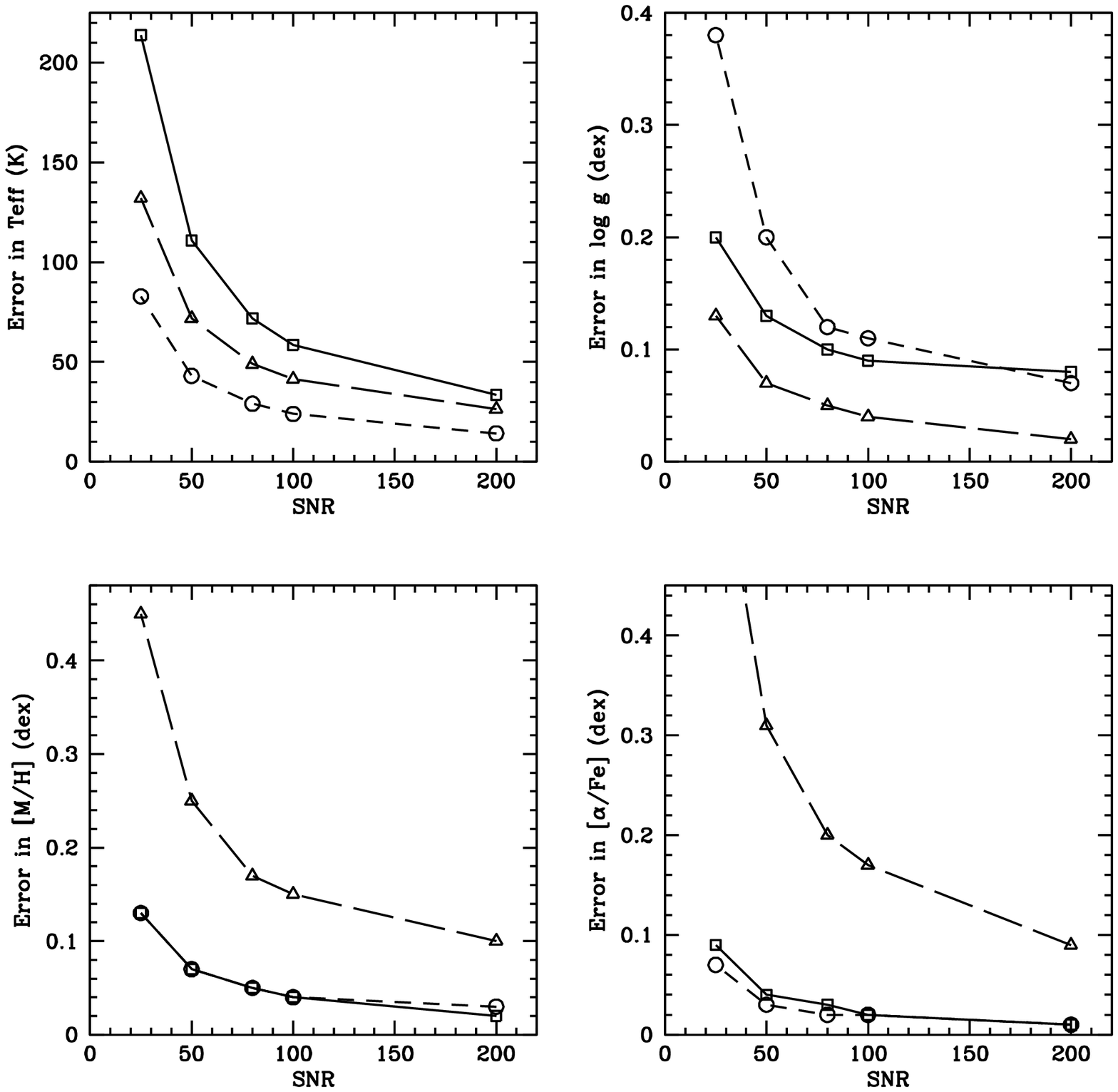}\\
\caption{Mean maximum errors in the recovered parameters as a function of the S/N, for synthetic 
continuum normalized spectra in the Gaia/RVS domain. The symbols are the same of Fig. 2}
\label{fig1}
\end{figure*}

The second step is the use of the local {\it B}$^{~l}_\theta$($\lambda$) functions
to derive the final stellar parameters. Once a centre of the subgrid has been chosen 
(e.g. $T_{\rm{eff}} = 5750$ K, log~$g = 4.5$ dex, [M/H]$ = 0.0$ dex, 
[$\alpha$/Fe]$ = 0.0$ dex, for solar type stars) the computed  {\it B}$_\theta$($\lambda$)
functions, can be used to analyse object spectra whose atmospheric parameters
and abundances are estimated to be near or inside the range covered by the subgrid. 
This includes any set of parameters, not necessary with an equivalent among the
points of the grid (e.g. the Sun: $T_{\rm{eff}} = 5777$ K, log~$g = 4.44$ dex, [M/H]$ = 0.0$ dex, 
[$\alpha$/Fe]$ = 0.0$ dex).

Figure~2 presents the mean maximum errors (mean bias plus mean standard
deviation) in the recovered parameters 
(effective temperature, gravity, global metallicity and [$\alpha$/Fe] content) 
as a function of the S/N, for the considered synthetic 
flux calibrated spectra in the Gaia/RVS domain. The three different lines on each panel correspond to
above defined metal rich cool dwarfs (solid line and squares), intermediate metallicity cool giants
(short dashed line and circles), metal poor hot subgiants (long dashed
line and triangles). In the following, the results for the three types of stars are discussed.

\subsubsection[]{Metal rich cool dwarfs}

The results indicate very small mean error (bias) values  at any S/N ratio: less than 
$\sim$ 10~K for the effective temperature, less than $\sim$ 0.03 dex for
the gravity, less than $\sim$  0.03~dex for [M/H]  and [$\alpha$/Fe]. The total maximum
error, as plotted on Fig. 2, includes, in addition, the mean standard deviation at each
S/N value. For S/N$\ge$50 it is smaller than $\sim$ 90~K, 0.14~dex, 0.05~dex
and 0.04~dex for $T_{\rm{eff}}$, 
log~$g$, [M/H] and [$\alpha$/Fe] respectively. At lower signal to noise values, the reduced
quantity of information makes the standard deviation increase, although the total errors
remain at acceptable levels down to S/N$=$25 ($\sim$ 173~K, 0.25~dex, 0.1~dex and 
0.06~dex respectively).


%
 
\subsubsection[]{Intermediate-metallicity cool giants}

As in the previous case, no significant biases were found in the derived parameters. 
The more remarkable difference, with respect to solar type stars, is the higher
accuracy on the  $T_{\rm{eff}}$ determination. Basically, this is due to the fact 
that the {\it B}$_{T_{\rm eff}}$($\lambda$) function has additional temperature indicators
(mainly molecular lines) and that its amplitude variations  
are larger by more than a factor of two with respect to the solar case (more numerous 
spectral lines are more sensitive to the
effective temperature). On the other hand, although still very well determined, the errors in the other 
parameters reflect a slightly smaller sensitivity of the calcium lines and the Pa 14 line to log~$g$,
and the smaller metal content of the analysed stars (with spectra having a smaller number and weaker
available lines).

\subsubsection[]{Very metal-poor hot subgiants}

The analysis of this kind of stars is worsened by the huge
decrease in the number and the intensity of metallic lines, leading to higher
errors in the global metallicity and the [$\alpha$/Fe] content. This is probably
the most challenging case for a chemical analysis, in general, due to the lack of information
(nevertheless, [M/H] and [$\alpha$/Fe] content are still well determined when
S/N$>$50). In contrast,
the Paschen lines are strengthened at this temperature and they allow a very
accurate determination of the gravity and satisfactory results for the effective temperature.

\subsection[]{Application to continuum normalized spectra}

This section takes into account the more common situation of non flux calibrated
spectra for which the continuum level is fitted by a polynomium to which the
whole spectrum is then normalized. 
To this purpose, a new grid of continuum normalized synthetic spectra was computed.
First of all, the MATISSE algorithm was
applied to normalized synthetic spectra in order to study the effects, on the performancies of the algorithm, 
of the loss of the information provided by the absolute stellar flux. 
Lastly, the observed continuum normalized spectra of two classical stars (the Sun and Arcturus) are
analysed to exemplify the application to observed data.

\subsubsection[]{Synthetic spectra}

A data base of a hundred continuum normalized synthetic spectra,
with the same randomly chosen parameters selected for the tests in Sect. 3.3,  
were treated. Again, 5 different S/N values and 1000 tests per signal to noise
ratio were performed for each spectrum. 

Figure~3 presents the corresponding mean maximum errors in the recovered parameters
as a function of the S/N, with the same division as Fig.~2 on 
metal rich cool dwarfs, intermediate metallicity cool giants, metal poor hot subgiants.

First, we can note that the performances for the global metallicity and the [$\alpha$/Fe] content are
not influenced by the absence of flux calibration. This implies the possibility of
deriving the [M/H] and the $\alpha$-element abundance with an accuracy better than 0.1~dex
for most types of stars, even with rather poor quality spectra.

On the other hand, the loss of the information about the effective temperature contained 
in the stellar flux (also revealed by the slope of a {\it B}$_{T_{\rm eff}}$($\lambda$) function
for flux calibrated spectra, like that of Fig. 1) induces a small increment of the
error (the standard deviation, not the bias) as the signal to noise decreases. Slight
error increments are also observed for the gravity determination of cool dwarfs and giants.
In any case, both temperature and gravity are still well determined for continuum normalized
spectra.

\subsubsection[]{Observed spectra} 
 
We present in the following two examples of the usage of MATISSE for the analysis 
of observed spectra, in order to justify and explain its application to  real
data treatment. As mentioned before, the analysis of observed spectra  suffers from various effects that
blur the extraction of the available information on a particular parameter:
approximations in the computation of theoretical spectra, 
errors in the wavelength calibration, treatment of the line broadening, etc... 

In order to check and illustrate the results of the algorithm
under those possible complications, two stars with well known atmospheric
parameters and chemical abundances have been considered: the Sun and Arcturus ($\alpha$ Boo).
In addition, they match two of the cases studied above through the tests with synthetic
spectra: metal rich cool dwarfs and intermediate metallicity cool giants.
The observed spectra have been taken from Hinkle et al. (2000) and their resolution
was degraded to the Gaia/RVS one, with the same sampling as the synthetic spectra
descrived in Section 3.1. 

As already mentioned in Section 3.1, the validity of the considered line 
data and the classical assumptions used (LTE, hydrostatic) have not been checked here
for all types of stars. Hence, biases on the derived parameters of
observed stars are expected to
be found. They can eventually be confirmed by a comparison between the results obtained for an observed
spectrum, and those for a synthetic one with the same atmospheric parameters (without biases).

In the case of the Sun ($T_{\rm{eff}} = 5777$~K, 
log~$g = 4.44$~dex, [M/H]$ = 0.0$~dex, [$\alpha$/Fe]$ = 0.0$~dex), 
the {\it B}$^{\rm{o}}_{T_{\rm eff}}$($\lambda$) funtions provided a first
parameter estimation with errors of 400~K, 0.34~dex and 1.2~dex in
$T_{\rm{eff}}$, log~$g$ and [M/H] respectively. That allowed to converge,
using the local {\it B}$^{~l}_{T_{\rm eff}}$($\lambda$) functions
to the following solution:
$T_{\rm{eff}} = 5758$~K, log~$g = 4.33$~dex, [M/H]$ = 0.07$~dex, 
[$\alpha$/Fe]$ = -0.05$~dex. 

%

For the Arcturus spectrum, the errors in the first parameter evaluation were
260~K, 0.20~dex and 0.9~dex in $T_{\rm{eff}}$, log~$g$ and [M/H]
respectively. The final 
derived values for Arcturus were: $T_{\rm{eff}} = 4277$ ~K, 
log~$g = 1.7$~dex, [M/H]$ = -0.40$~dex, [$\alpha$/Fe]$ = 0.15$~dex. These values can be compared to
those published by Peterson et al. (1993): $T_{\rm{eff}} = 4300$~K $\pm$~30~K, 
log~$g = 1.5$ $\pm$~0.15~dex, [M/H]$ = -0.5$ $\pm$~0.1~dex, and [$\alpha$/Fe] ranging 
from 0.3 $\pm$~0.1~dex to 0.4 $\pm$~0.1~dex, depending on the considered $\alpha$-element.

Finally, as already mentioned, the small differences (bias) between the derived parameters for the
Sun and Arcturus and those in the literature could arise from inadequations of
the theoretical spectra, mainly implemented for the test of the algorithm.
To verify this possibility, synthetic continuum-normalized spectra, with
atmospheric parameters equal to those of the Sun and Arcturus,
were calculated and treated with the same {\it B}$_\theta$($\lambda$)
functions as the observed ones. The errors (biases) in
the results were actually considerably smaller: for the synthetic Sun, $\Delta T_{\rm{eff}}$ = 6~K, 
$\Delta$log~$g$ = 0.02~dex, $\Delta$[M/H] = 0.01~dex and $\Delta$[$\alpha$/Fe] = 0.01~dex; 
for the synthetic Arcturus, 1~K, 0.01~dex, 0.004~dex and 0.006~dex, respectively. 
As a consequence, we conclude that the small biases in the results presented here are most probably 
coming from imperfections in the physical assumptions and the line data used for the theoretical 
spectra computation.
The suitability of the MATISSE algorithm to the analysis of
observed spectra is therefore confirmed.

\section[]{Discussion and conclusions} 
 
The illustrated results corroborate the efficiency of the automated MATISSE
algorithm as a spectral analysis and classification tool. In particular, 
the capabilities of MATISSE to accurately derive stellar atmospheric parameters
and chemical abundances (as the [$\alpha$/Fe] ratio) have been described.
The method gives rapid, compelling and stable results, without biases, 
even for moderate to low signal to noise spectra and flux normalized data.

The accuracy values presented here correspond
to a red spectral domain, that of the Gaia/RVS, where the number of metallic
lines, and therefore the quantity of information, is smaller than at bluer wavelengths. 
Even better results could indeed be achieved in different spectral domains or by increasing 
the wavelength range of analysis or the spectral resolution. The stable performances for
stars in different regions of the Hertzsprung-Russel diagram and the applications to
chemical abundance determinations make of MATISSE a powerful tool for
the study of stellar populations.
In particular, the accuracies attained on chemical abundances are better than about
0.1~dex (except for very low-metallicity hot stars with too few available lines), 
as required  to  constrain Galactic formation and evolution
models (see for  example, Prantzos, 2003; Robin et al.,  2000). 
Finally, no border effects (error increasing for object spectra with
parameters near the limits of the considered spectra grid), 
with respect to the synthetic spectra grid,
are disturbing the results, at variance with what is usually the case 
for minimum distances methods.

This algorithm represents a new effort in automated spectral analysis, different from 
the already existent automated classification techniques. 
In 1985, Cayrel de Strobel claimed, from theoretical considerations, 
that metallicity can be determined from high resolution spectra,
of known effective temperature and gravity,
with an accuracy of $\pm$0.07~dex at S/N$=$250 and of $\pm$0.2~dex at S/N$=$50.
Later on, Jones et al. (1996) derived [Fe/H] values for G stars, through the
use of spectroscopic indices at 1\AA~ resolution in the range 4000-5000\AA. They achieved
errors of $\pm$0.2~dex at low S/N (10-20).  Since then, the main efforts
on automated stellar parameter determination could be divided in three categories: minimum distances
techniques, genetic algorithms and neural networks. In this context, the method presented
here opens an unexplored avenue.

The minimum distances technique was
applied by Katz et al. (1998), using a template grid of flux calibrated 
observed spectra of FGK stars with solar composition,
in a much larger wavelength range and at a higher spectral resolution than those presented here
(a 2900\AA ~region around 5300\AA, with R$=$42\,000).
They estimated the internal accuracy of their method to be 86~K, 0.28~dex and 0.16~dex for
$T_{\rm{eff}}$, log~$g$ and [M/H] at S/N$=$100, that is, (except for $T_{\rm{eff}}$) much higher errors than those
achieved with MATISSE. 

Bonifacio \& Caffau (2003)  performed an automated abundance analysis of giants
in the Sagittarius dwarf-spheroidal galaxy using a $\chi^2$ technique. They
achieved an accuracy in the [Fe/H] determination of 0.18~dex, for spectra of stars with known 
$T_{\rm{eff}}$ and log~$g$, R$=$15\,000 and S/N$=$20. Some stability problems at low
metallicity are also reported.

Allende Prieto (2003) applied a genetic algorithm to determine stellar
atmospheric parameters from spectra, in a 150\AA ~region around H$\beta$, of
A- to K-type stars with S/N $\simeq$ 150. The estimated 1$\sigma$ errors for R$\simeq$5000 resolution,
 are 100~K, 0.3~dex, 0.1~dex for $T_{\rm{eff}}$, log~$g$ and [Fe/H], respectively.

A growing number
of works have been developed using artificial neural networks. 
Bailer-Jones (2000) reports, for very low resolution (R$=$60-100) and low signal to
noise (S/N$=$5-10) spectra, in a very large wavelength domain (a 7000\AA ~region, centered on
6500\AA) errors of 50-300~K in $T_{\rm{eff}}$, $\geq$0.2~dex in log~$g$ and 0.2~dex in
[M/H], across practically the whole HR diagram. 
 Snider et al. (2001) tested the application of artificial neural networks on
a sample of G- and F-type stars, with spectra in a $\sim$ 850 \AA ~domain in the blue, at
medium resolution (R$=$2000-4000). They attained accuracies of 135-150~K, 
0.25-0.30~dex and 0.15-0.20~dex  in effective temperature, gravity and [Fe/H], respectively.
Lastly, Willensen et al. (2005) employed feed forward neural networks, trained on synthetic spectra
in a 1800\AA ~region around 4700\AA ~(R$\simeq$1500-2400), to determine metallicities  of main-sequence 
turn-off, subgiant and red giant stars in two globular clusters. 
They reported uncertaines, for S/N$=$40-80, of 140-190~K in $T_{\rm{eff}}$,
0.3-0.4~dex in log~$g$ and 0.15-0.2~dex in [Fe/H].

At variance with genetic algorithms and artificial neural networks, the automated method
presented here exploits the possibility of easily selecting a parameter domain (the grid subregion)
where the spectrum variations are practically linear.   
Summing up, the performances of the MATISSE algorithm are better or, at least,
comparable to previous results on automated analysis, for different
spectral domains and resolutions. They are also competitive with the results
obtained by fine analysis of high-resolution spectra. In addition, [$\alpha$/Fe]
automated derivations are presented for the first time, with very positive results. 
The MATISSE algorithm could be applied also to a classification based on
photometric indices or spectra of non resolved galaxies.

On another hand, regarding the computational time needed to derive the
atmospheric parameters of an unknown star, the efficiency of the MATISSE 
algorithm is very high. Indeed, once the $B_\theta(\lambda)$ have been
derived for different locations of the HR diagram (or subgrids), 
the stellar parameters are almost instantaneously derived from
Eq.~1. 
For each parameter, only a multiplication of two vectors which dimension
is equal to the number of sampling elements in the spectra
has to be carried out. As a consequence, the atmospheric parameters and the [$\alpha$/Fe] content of the 
whole Gaia/RVS spectra database could be evaluated in about
1~hour, with only one present day processor running at 0.5GFLOP/sec.
On the other hand, the derivation of the $B_\theta(\lambda)$ functions
(covariance matrix calculation between the spectra of the grid and
solution of Eq.~6) takes less than 1 month of computational time with one
present day technology 0.5GPLOP/s processor, for the case of the grid presented in Section 3.1.
This time can be exactly divided by the number of available processors,
as the different  $B_\theta(\lambda)$ functions are calculated independently
(e.g. $\sim$ 1~week of computing time with 4 processors).
%

We can finally conclude that MATISSE is  particularly adapted to the analysis
of large surveys with huge amount of spectra, as the Gaia mission.
Further applications of the algorithm, as the measurement
of individual stellar chemical abundances, stellar rotation and microturbulence,
will be implemented in the future. This will allow to generalize and exploit, as much as
possible, the capabilities of the MATISSE algorithm on stellar spectra analysis.

\section*{Acknowledgments} 
 
We thank the MARCS collaboration for providing us the grid of
model atmospheres in advance of publication and B. Plez for his molecular line lists and tools 
for computing stellar spectra.
The VALD database was used for the atomic lines. J.C. Gazzano is acknowledged for his help with
Fig.~1 of this paper and we are grateful to D. Katz for discussions and advices regarding
the Gaia Radial Velocity Spectrograph. We acknowledge F. Th\'evenin  for his initial interest in
this work. We sincerely thank the
anonymous referee whose corrections and advices have sensitively improved this paper. 
ARB acknowledges the support of the European Space Agency.

 

\bsp 
 
\label{lastpage} 
 
\end{document}